\documentclass[fleqn,usenatbib]{mnras}

\usepackage{newtxtext,newtxmath}

\usepackage[T1]{fontenc}
\usepackage{ae,aecompl}

\usepackage{graphicx}	
\usepackage{amsmath}	
\usepackage{amssymb}	
\usepackage{times}
\usepackage{ulem}
\usepackage{xcolor}

\defcitealias{2017MNRAS.467.1984G}{GLM17}
\newcommand{\dd}{\mathrm{d}}

\usepackage{etoolbox}
\makeatletter
\patchcmd\@combinedblfloats{\box\@outputbox}{\unvbox\@outputbox}{}{\errmessage{\noexpand patch failed}}
\makeatother

\title{ {Self-Induced Dust Traps Around Snow Lines in Protoplanetary Discs}}

\author[Vericel \& Gonzalez]{
Arnaud Vericel,$^{1}$\thanks{E-mail: arnaud.vericel@univ-lyon1.fr}
Jean Fran\c cois Gonzalez $^{1}$
\\
$^{1}$Univ Lyon, Univ Claude Bernard Lyon 1, ENS de Lyon, CNRS, Centre de Recherche Astrophysique de Lyon UMR5574, F-69230, Saint-Genis-Laval, France\\
}

\date{Accepted 2019 December 4. Received 2019 November 26; in original form 2019 October 17}

\pubyear{2020}

\begin{document}
\label{firstpage}
\pagerange{\pageref{firstpage}--\pageref{lastpage}}
\maketitle
\begin{abstract}
Dust particles need to grow efficiently from micrometre sizes to thousands of kilometres to form planets. With the growth of millimetre to meter sizes being hindered by a number of barriers, the recent discovery that dust evolution is able to create ``self-induced'' dust traps shows promises. The condensation and sublimation of volatile species at certain locations, called snow lines, is also thought to be an important part of planet formation scenarios. Given that dust sticking properties change across a snow line, this raises the question: how do snow lines affect the self-induced dust trap formation mechanism? The question is particularly relevant with the multiple observations of the carbon monoxide (CO) snow line in protoplanetary discs, since its effect on dust growth and dynamics is yet to be understood. In this paper, we present the effects of snow lines in general on the formation of self-induced dust traps in a parameter study, then focus on the CO snow line. We find that for a range of parameters, a dust trap forms at the snow line where the dust accumulates and slowly grows, as found for the water snow line in previous work. We also find that, depending on the grains sticking properties on either side of the CO snow line, it could either be a starting or braking point for dust growth and drift. This could provide clues to understand the link between dust distributions and snow lines in protoplanetary disc observations.
\end{abstract}




\section{Introduction}
\label{intro}

Over the last few decades, our perception of planet formation has changed significantly because of the continuous improvement in the resolution of protoplanetary disc observations. Additionally, the growing number of exoplanet discovered each year indicates that planet formation is very common and diverse \citep{2012Natur.481..167C}, which also seems to be in agreement with the fascinating diversity of structures seen in recent observations of discs \citep{2018ApJ...869L..41A}. Understanding how to link these clues and building a consistent planet formation theory is the goal of both theoreticians and observers for years to come.

On the theoretical side, the core accretion paradigm struggles to explain how solids can reach sizes of several thousand kilometres \citep{1977MNRAS.180...57W}. In this scenario, dust particles co-evolve with the gas in protoplanetary discs \citep{1973NASSP.319..355W} and grow by coagulation with other dust particles \citep{1993prpl.conf.1061L,1997ApJ...480..647D,2005A&A...434..971D}. The gas is sensitive to its own pressure gradient, which makes it orbit the star at a sub-Keplerian velocity, while the dust orbits at a Keplerian velocity. This velocity difference causes a headwind on the dust, which removes angular momentum from the grains and induces a radial drift towards the star \citep{1977MNRAS.180...57W,1986Icar...67..375N}. The small dust grains are strongly coupled to the gas, while the large ones are strongly decoupled. In both cases the radial drift is slow. It is maximal for intermediate-sized grains, typically between the millimetre and the centimetre. As grains grow, they reach this maximal radial drift velocity and are rapidly accreted onto the star. This constitutes one of the obstacles for planet formation, first known as the ``meter-size barrier'' in the Minimum Mass Solar Nebula (MMSN) model \citep{1977MNRAS.180...57W,1981IAUS...93..113H} and later more generally as the ``radial-drift barrier'' \citep{2012A&A...537A..61L}.
Similarly to the dust radial drift, the relative velocity between dust particles becomes large at intermediate sizes \citep{WC1993}, which makes grains bounce or shatter rather than stick. These planet formation barriers have been referred to as the ``bouncing'' and ``fragmentation'' barriers \citep{2008ARA&A..46...21B,2010A&A...513A..57Z}. Our understanding of planet formation is tied to this intermediate size regime, where the dust has to continue to grow further in order to remain in the disc and eventually form planets.

To overcome the radial-drift barrier and avoid accretion, dust grains need to grow very rapidly to experience the fastest radial drift velocity regime for the shortest amount of time. Alternatively, as the dust-to-gas ratio increases, the collective effects of the dust onto the gas become stronger and can lead to slower radial drift \citep{1986Icar...67..375N}. Finally, one can also trap dust into a local pressure maximum, thus halting its drift \citep{2005MNRAS.362.1015H}. The ensuing particle concentration in a dust trap has the additional benefit of reducing the dust relative velocity, thus helping overcoming the bouncing and fragmentation barriers. Several mechanisms have been proposed to create pressure bumps, such as dead zones \citep{2007ApJ...664L..55K,2010A&A...515A..70D}, vortices \citep{1995A&A...295L...1B,2014ApJ...795...53Z} or planet gaps \citep[]{2004A&A...425L...9P,2007A&A...474.1037F,2012MNRAS.423.1450A,2015P&SS..116...48G,2017MNRAS.469.1932D}.
While these mechanisms require special conditions in discs, \citet[][hereafter GLM17]{2017MNRAS.467.1984G} showed that the back-reaction on the gas of growing and fragmenting dust grains in discs with large scale gradients is a powerful way to naturally form dust traps and therefore overcome planet formation barriers. The mechanism has been called ``self-induced dust trap'' because of its ability to form on its own. This paper will focus on this mechanism of dust trapping.

Across the range of pressures and temperatures in protoplanetary discs, some material can experience a transformation between gaseous and solid states because of the wide range of pressures and temperatures. This leads to the existence of a condensation front, where a given species condenses (or sublimates) at the surface of grains. This front is called a ``snow line'' \citep{2006ApJ...640.1115L,2011ApJ...735...15G} and is essential in understanding planetesimal compositions \citep{2016Natur.535..237M}. 
Snow lines are particularly interesting for dust growth and drift because they affect the grain properties. If a volatile species is abundant enough, the snow line can play an important role in grain evolution, affecting growth and radial drift \citep{2008ARA&A..46...21B}. When grains cross a snow line, one also expects a diffusion of freshly sublimated gas both inwards and outwards \citep{2006Icar..181..178C}. Gas moving outwards and crossing the snow line condenses onto the surface of grains, increasing the dust surface density \citep[][for the water snowline specifically]{2016ApJ...828L...2A,2017A&A...602A..21S,2017A&A...608A..92D}. Both of these processes can help radially concentrate dust.

Historically, the water snow line has been the most studied due to its proximity to the terrestrial planet forming region, its important implications for the composition of planets and their atmospheres, and our extensive knowledge of the behaviour of water ice \citep{0004-637X-818-1-22}. However, the water snow line, located within the first astronomical units of the disc \citep{2004M&PS...39.1859P}, is difficult to resolve in current observations. As a comparison, the DSHARP survey \citep{2018ApJ...869L..41A} has a resolution of $\sim$5~au at 150~pc -- at least one order of magnitude too high to observe the water snow line around Classical T-Tauri Stars (CTTS).
Nevertheless, the water snow line is probably the most meaningful for the grains, since it separates dry silicate cores from wet icy aggregates. The collisional energy necessary to break-up icy aggregates is much higher than for bare silicates \citep{2007DPS....39.3204T,2008ARA&A..46...21B,2019ApJ...874...60S}, and as a consequence the collisional growth of grains interior to the water snow line is severely hindered \citep{2018ApJ...868..118H} compared to the outside. It is worth mentioning that recent experimental work from \citet{2019ApJ...873...58M} indicate that icy aggregates behaviour could also be dependent on temperature and that it could fragment more easily in colder regions of the disc. Water is also one of the most abundant volatile species in discs, leading to a substantial diffusion at the snow line location \citep{2016A&A...596L...3I,2017A&A...602A..21S}. The combination of these effects can trigger the streaming instability and form planetesimals in immediate proximity to the water snow line for weakly turbulent discs \citep{2017A&A...608A..92D}. Observations show that the water snow line also potentially affects the occurence of giant planets \citep{2019ApJ...874...81F}. The recent discovery of a super-Earth orbiting Barnard's Star b at the location of the water snow line \citep{2018Natur.563..365R} provides further evidence of the importance of the ice line in planet formation.

Alternatively, colder snow lines have been observed in several discs, in particular the carbon monoxide (CO) snow line, which is located at much larger distances -- a few tens to a hundred au \citep{2013A&A...557A.132M,2015ApJ...813..128Q,2016A&A...588A.112G,2017ApJ...838...97M,2018IAUS..332...88V,2018A&A...609A..47P}. Unfortunately, its effect on grains is poorly understood. While the water snow line is expected to be linked to dust structures, it is unclear if we can state the same for CO. For example, \citet{2016A&A...588A.112G} probed the dust distribution around the CO snow line in HD 163296, measured by \citet{2015ApJ...813..128Q} at $\sim 90$~au using multiple tracers. They highlighted an apparent lack of large grains exterior to the snow line. As they point out, more theoretical work needs to be done to better interpret these observations.

In order to improve our understanding of planet formation, we must better understand the impact of snow lines on dust evolution. This paper aims to contribute to this goal: by performing global numerical simulations of discs over a wide parameter space, we provide an overview of the possible effects snow lines can have on self-induced dust trap formation and evolution. We also investigate the role of the CO snow line to provide insights into dust structures for both observers and theoreticians. We present our growth and fragmentation model in Section~\ref{method} and the main results in Section~\ref{res}. We discuss our results in Section~\ref{discu} and finally conclude in Section~\ref{conclu}.

\section{Methods}
\label{method}

\subsection{Gas and dust dynamics}
\label{gd-dyn}

Taking into account the dust back-reaction, the stationary solutions of the equations of motion for the gas and dust radial velocities are \citep{2017ApJ...844..142K, 2018MNRAS.479.4187D}

\begin{equation}
v_\mathrm{g,r} = - \dfrac{\varepsilon \mathrm{St}}{(1+\varepsilon)^2+\mathrm{St}^2}v_\mathrm{drift} + \dfrac{1+ \varepsilon+ \mathrm{St}^2}{(1+\varepsilon)^2 + \mathrm{St}^2}v_\mathrm{visc},
\label{vgr}
\end{equation}

\begin{equation}
v_\mathrm{d,r} =  \dfrac{\mathrm{St}}{(1+\varepsilon)^2+\mathrm{St}^2}v_\mathrm{drift} + \dfrac{1+\varepsilon}{(1+\varepsilon)^2 + \mathrm{St}^2}v_\mathrm{visc},
\label{vdr}
\end{equation}
where $\mathrm{St}$ is the Stokes number of dust particles, $\varepsilon$ is the dust-to-gas ratio,
\begin{equation}
v_\mathrm{drift} = \left(\dfrac{H}{r}\right)^2 \dfrac{\partial \log P_\mathrm{g}}{\partial \log r} v_\mathrm{k}
\end{equation}
is the optimal drift velocity caused by the drag \citep{1986Icar...67..375N}, $v_\mathrm{k}$ is the Keplerian velocity and
\begin{equation}
v_\mathrm{visc} = \dfrac{\dfrac{\partial}{\partial r}\left(\rho_\mathrm{g}\nu r^3\dfrac{\partial \Omega_\mathrm{k}}{\partial r}\right)}{r\rho_\mathrm{g}\dfrac{\partial}{\partial r}(r^2\Omega_\mathrm{k})}
\end{equation}
is the viscous velocity \citep{1974MNRAS.168..603L}.
The drift and viscous velocities being negative, the dust drifts towards to star. 
For the gas, $v_\mathrm{visc}$ dominates $v_\mathrm{drift}$ for small dust-to-gas ratios and small sizes, leading to the gas accretion onto the star.
However, when the dust-to-gas ratio increases and the Stokes number is close to unity, the positive first term on the right hand side of equation \ref{vgr} increases and has the effect of decreasing the inwards gas radial velocity. For sufficiently large dust-to-gas ratios, this collective effect term is also able to dominate the viscosity-induced velocity and revert the motion of gas towards the outer disc.
While these stationary solutions are not directly used in our simulations (the equations of motion are instead solved directly, see Sect.~\ref{hydro}), they are useful to understand which mechanism is responsible for creating a pressure maximum somewhere in the disc other than at the inner edge. This mechanism is at the center of the self-induced dust trap mechanism.

\subsection{Hydrodynamical simulations}
\label{hydro}
To simulate protoplanetary disc evolution, we use our 3D, two-phase (gas+dust), Smoothed Particles Hydrodynamics (SPH) code \citep{2005A&A...443..185B}. It computes the forces acting on each SPH particle and solves their equation of motion. Gas-dust aerodynamic coupling is incorporated as described in \citet{1997JCoPh.138..801M} taking into account the backreaction of the dust onto the gas. Table~\ref{effects} shows a more detailed overview of what is included in our numerical setup.

We model a $0.01$~M$_\odot$ disc orbiting a $1$~M$_\odot$ star. Initially, we set 200,000 particles representing the gas disc with a power-law surface density $\Sigma = \Sigma_0 (r/r_0)^{-\it{p}}$. The temperature structure is vertically isothermal and also follows a radial power-law, $T = T_0 (r/r_0)^{-\it{q}}$. We allow the gas particles to evolve for 8 orbits at 100 au (8000~yr) to reach steady state, and then inject an equal number of dust particles, such that the initial dust-to-gas ratio $\epsilon$ is uniform and equal to $0.01$. Grains have an initial size of 10~$\mu$m and are able to grow or fragment as detailed in Section~\ref{growth}. Both gas and dust particles are set between $r_\mathrm{in}$ and $r_\mathrm{out}$ and are removed from the simulation if they cross $r_\mathrm{esc}$. For further information on the code and setup, we refer the reader to \citetalias{2017MNRAS.467.1984G}. The disc model parameters are given in Table~\ref{steep}. Resolution studies with our code have shown convergence with fewer particles \citep{2005A&A...443..185B,2007A&A...474.1037F,2019MNRAS.490.4428P}. Additionally, the resolution criterion proposed by \citet{2012MNRAS.420.2345L} is met in our simulations.

\begin{table}
\centering
\begin{tabular}{ccccccc}
\hline
$p$ & $q$ & $\Sigma_0$ & $T_0$ & $r_{\mathrm{in}}$ & $r_{\mathrm{out}}$ & $r_{\mathrm{esc}}$ \\
 & & $\left[\mathrm{kg}\,\mathrm{m}^{-2}\right]$ & $\left[\mathrm{K}\right]$ & $\left[\mathrm{au}\right]$ & $\left[\mathrm{au}\right]$ & $\left[\mathrm{au}\right]$\\
\hline
$1$ & $1/2$ & $487.74$ & $200$ & $10$ & $300$ & $400$ \\

\hline
\end{tabular}
\caption{The disc model used in our simulations, with $r_0 = 1$ $\mathrm{au}$.}
\label{steep}
\end{table}

The coupling between gas and dust is represented by the Stokes number, which is the ratio between the stopping time $t_\mathrm{s}$ of a dust particle and its Keplerian orbital time $t_\mathrm{k}$: $\mathrm{St} = t_\mathrm{s} / t_\mathrm{k}$. In our simulations, the gas spatial density is low and the grains have smaller sizes than the gas mean free path: $s<9/4\lambda_\mathrm{g}$. We subsequently treat the dust dynamics in the Epstein regime \citep{1924PhRv...23..710E}, where the Stokes number can be expressed as
\begin{equation}
\mathrm{St} = \dfrac{\Omega_\mathrm{k}\rho_{\mathrm{s}}s}{\rho_\mathrm{g}c_\mathrm{s}},
\label{stokesnumb}
\end{equation}
where $\Omega_\mathrm{k}$ is the Keplerian frequency, $\rho_\mathrm{s}$ the dust intrinsic density, $s$ the grain size, $\rho_\mathrm{g}$ the volume density of the gas phase and $c_\mathrm{s}$ the gas sound speed defined as

\begin{equation}
c_\mathrm{s} = \sqrt{\dfrac{k_\mathrm{B}T}{\mu m_\mathrm{H}}} = c_\mathrm{s,0} \left(\dfrac{r}{r_0}\right)^{-\frac{q}{2}},
\label{csoundeq}
\end{equation}
where $k_\mathrm{B}$ is the Boltzmann constant, $\mu$ the mean molecular weight and $m_\mathrm{H}$ the mass of the hydrogen atom.

\begin{table}
\centering
\begin{tabular}{lc}
Mechanism & Included \\
\hline
Growth & Yes \\
Fragmentation & Yes \\
Back-reaction & Yes \\
Snow line (different $V_\mathrm{frags})$ & Yes \\
Evaporation/condensation & No \\
Self-gravity & No \\
\hline
\end{tabular}
\label{effects}
\caption{Summary of the mechanisms included in our model.}
\end{table}

\subsection{Growth and fragmentation models}
\label{growth}

The implementation of grain growth follows the prescription of \citet{2008A&A...487..265L}. We use a mono-disperse approximation for the grains, i.e. we assume that for each dust SPH particle, the size distribution is strongly peaked around a mean value. The relative motion between the grains allows them to grow if their relative velocity is lower than a fragmentation threshold, $V_\mathrm{frag}$. The turbulent relative velocity between dust grains is given by \citet{1997A&A...319.1007S}:

\begin{equation}
V_\mathrm{rel} = \sqrt{2^{3/2}\mathrm{Ro}\alpha} \dfrac{\sqrt{1-\mathrm{Sc}}}{\mathrm{Sc}}c_\mathrm{s},
\label{Vrel}
\end{equation}
where $\mathrm{Ro}$ is the Rossby number, considered constant and equal to $3$ and $\alpha$ is the viscosity parameter from \citet{1973A&A....24..337S} set to $10^{-2}$. $\mathrm{Sc}$ is the Schmidt number, given by

\begin{equation}
\mathrm{Sc} = \left(1 + \mathrm{St}\right) \sqrt{1 + \dfrac{\Delta \boldsymbol{v}^2}{V_\mathrm{t}^2}},
\label{Schmidt}
\end{equation}
where $V_\mathrm{t} = \sqrt{2^{1/2}\mathrm{Ro}\alpha}\,c_\mathrm{s}$ is the turbulent velocity and $\Delta \boldsymbol{v} = \boldsymbol{v}_\mathrm{d} - \boldsymbol{v}_\mathrm{g}$ is the differential velocity between the dust and gas phases. We only consider the turbulent relative velocity term in our model, which we verify to be dominant compared to the radial drift or brownian components. Moreover, our model being mono-disperse, we do not model interactions between grains of different sizes and consequently have a differential radial drift velocity that is null within this approximation. This is one of the caveats of our model as we discuss in Section~\ref{sim-obs}.

When a dust particle grows (i.e. if $V_\mathrm{rel} < V_\mathrm{frag}$), it doubles its mass during the collision time: $\mathrm{d}m_\mathrm{d}/\mathrm{d}t = m_\mathrm{d}/\tau_\mathrm{col}$, which translates in size to
\begin{equation}
\dfrac{\mathrm{d}s}{\mathrm{d}t} = \dfrac{\rho_\mathrm{d}}{\rho_\mathrm{s}}V_{\mathrm{rel}},
\label{dsdt}
\end{equation}
where $\rho_\mathrm{d}$ is the volume density of the dust phase.
When the grains fragment (i.e. when $V_\mathrm{rel} > V_\mathrm{frag}$), their size evolve following \citet{2015P&SS..116...48G}:
\begin{equation}
\dfrac{\mathrm{d}s}{\mathrm{d}t} = - \dfrac{\rho_\mathrm{d}}{\rho_\mathrm{s}}V_{\mathrm{rel}},
\label{wSt}
\end{equation}
which, similar to the growth case, means that the initial grain loses most of its mass during the collision. This corresponds to a catastrophic fragmentation event.

\subsection{Snow lines as discontinuities in fragmentation threshold}
\label{snow}

\begin{figure}
\centering
\resizebox{\hsize}{!}{
\includegraphics[width=\columnwidth]{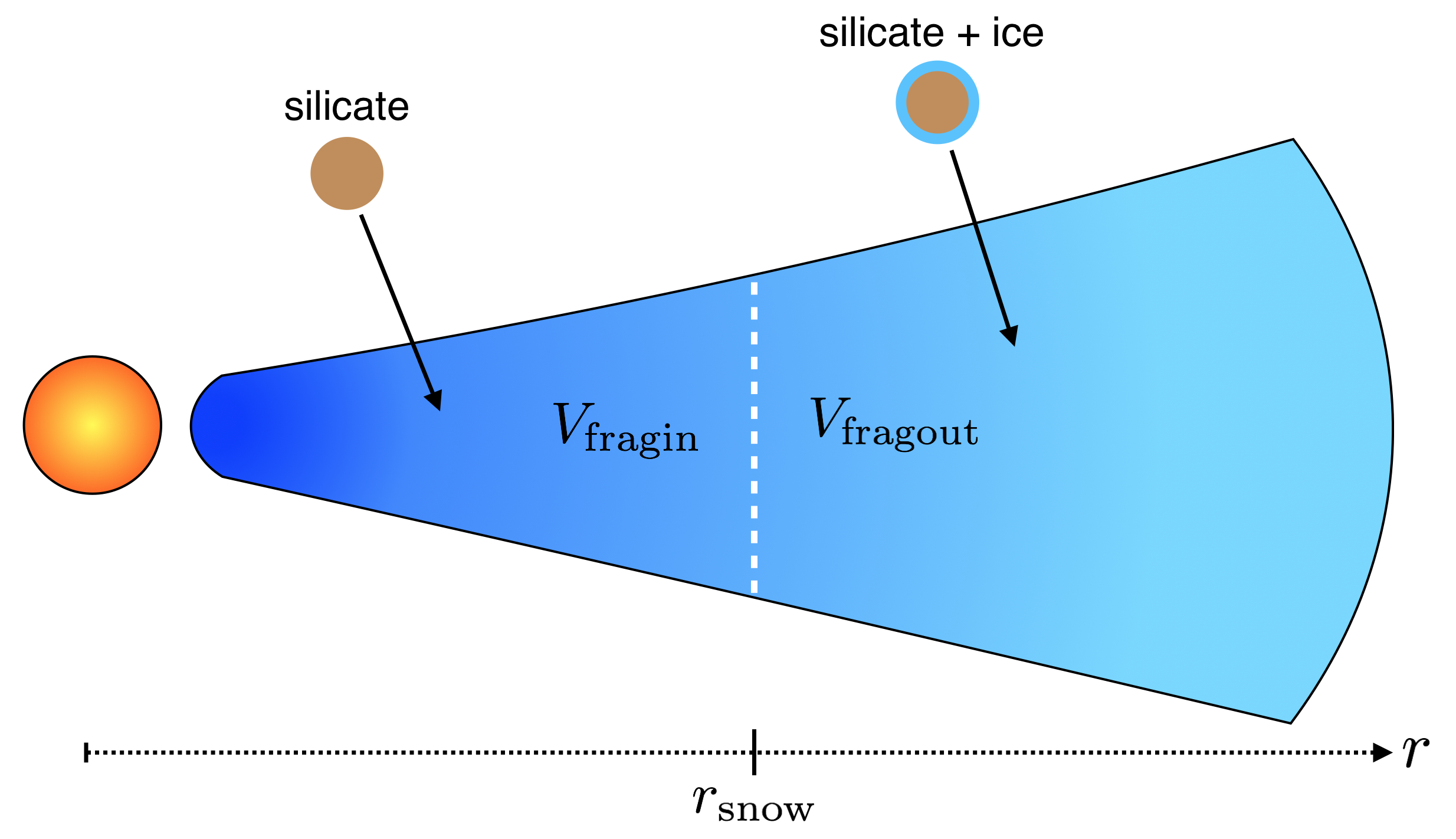}
}
\caption{Schematic of a protoplanetary disc seen edge-on. The temperature and pressure drop as the radial distance increases, which allows some icy volatile species to condense onto the surface of solid grains (here represented by silicates). The condensation (or sublimation) front is called snow line and is represented here by a separation at a given distance $r_\mathrm{snow}$ (white dashed line). The dust sticking properties in each zone are represented by $V_\mathrm{fragin}$ and $V_\mathrm{fragout}$.}
\label{snowlinescheme}
\end{figure}

The fragmentation velocity $V_\mathrm{frag}$ has been experimentally studied for silicates, water ice and $\mathrm{CO}_2$ ice \citep{2008ARA&A..46...21B,2009ASPC..414..347W,2010A&A...513A..56G,2014ApJ...783L..36Y,2016ApJ...827...63M,2019ApJ...873...58M}  and has been shown to be dependent on the dust composition. Fragmentation velocities are in the range between 1~m\,s$^{-1}$ (silicates) and several tens of m\,s$^{-1}$ (icy aggregates, see \citealt{2015P&SS..116...48G} for a detailed review). In discs, the pressure and temperature span several orders of magnitude, which causes volatile material to change state at certain locations. To take this into account, we incorporate in our simulation a snow line (see Fig.~\ref{snowlinescheme}). To represent the effect of that snow line, we adopt different values of the fragmentation velocity such that $V_\mathrm{fragin}$ corresponds to the fragmentation threshold interior to the snow line and $V_\mathrm{fragout}$ to the exterior threshold. This change in $V_\mathrm{frag}$ mimics the change in grain surface composition, meaning that the smaller $V_\mathrm{frag}$ is, the weaker the corresponding grain will be regarding fragmentation. 

We define the position of the snow line, $r_\mathrm{snow}$, either as the location where the temperature is equal to the sublimation temperature $T_\mathrm{subl}$:
\begin{equation}
r_\mathrm{snow} = r_0 \left(\dfrac{T_0}{T_\mathrm{subl}}\right)^\frac{1}{q},
\label{rsnow}
\end{equation}
or arbitrarily.
When the location is arbitrary, we use it to study the generic effect of a hypothetical snow line on dust evolution. The only physical snow line that we model in this paper is that of carbon monoxide ($\mathrm{CO}$), which is the only iceline (to date) to be consistently observed in discs due to its large radial distance from the star \citep{2013A&A...557A.132M}. Icy carbon monoxide aggregates could be much weaker than water ice aggregates and behave as silicates as proposed by \citet{2017ApJ...845...68P} for $\mathrm{CO_2}$. This behaviour could be due to the stronger chemical bond between hydrogen and oxygen atoms than those between carbon and oxygen atoms. However, at several tens of au from the star, dust is not only composed of CO, but is rather a mixture of all the icy species at this location (mainly $\mathrm{NH_3}$, $\mathrm{H_2O}$, $\mathrm{CO_2}$ and $\mathrm{CO}$). As the fragmentation behaviour for carbon monoxide aggregates is uncertain (increasing or decreasing $V_\mathrm{frag}$ when sublimating), we chose to consider both cases and explore the parameter space ($V_\mathrm{fragin}$  {and} $V_\mathrm{fragout}$). For more generic simulations, we also varied the snow line location $r_\mathrm{snow}$. The simulations ran for this paper are shown in table \ref{simu-table}. The nomenclature is S followed by the value of $r_\mathrm{snow}$ in au, then V followed by the values of $V_\mathrm{fragin}$ and $V_\mathrm{fragout}$ in m\,s$^{-1}$ . To match experimental studies as well as observations, we kept the fragmentation velocities between 1 and 15~m\,s$^{-1}$  and the snow lines between 20 and 200 au.

\begin{table}
\centering
\begin{tabular}{lcccc}
\hline
Label & $r_\mathrm{snow} $ & $V_\mathrm{fragin} $ & $V_\mathrm{fragout}$ & $V_\mathrm{fragin}/V_\mathrm{fragout}$ \\
 & $\mathrm{[au]}$ & $[\mathrm{m\,s^{-1}}]$ & $[\mathrm{m\,s^{-1}}]$ \\
\hline
S15V5-15 & 15 & 5 & 15 & 1/3 \\
S20V5-15 & 20 & 5 & 15 & 1/3 \\
S30V5-15 & 30 & 5 & 15 & 1/3 \\
S40V5-15 & 40 & 5 & 15 & 1/3 \\
S50V1-15 & 50 & 1 & 15 & 1/15 \\
S50V3-15 & 50 & 3 & 15 & 1/5 \\
S50V5-15 & 50 & 5 & 15 & 1/3 \\
S50V10-15 & 50 & 10 & 15 & 2/3 \\
S50V15-10 & 50 & 13 & 10 & 1.5 \\
S50V15-5 & 50 & 15 & 5 & 3 \\
SS75V5-15 & 75 & 5 & 15 & 1/3 \\
S100V1-15 & 100 & 1 & 15 & 1/15 \\
S100V3-15 & 100 & 3 & 15 & 1/5 \\
S100V5-15 & 100 & 5 & 15 & 1/3 \\
S100V10-15 & 100 & 10 & 15 & 2/3 \\
S100V15-5 & 100 & 5 & 5 & 3 \\
S100V15-10 & 100 & 1 & 15 & 1.5 \\
S150V3-15 & 150 & 3 & 15 & 1/5 \\
S150V5-15 & 150 & 5 & 15 & 1/3 \\
S150V12-15 & 150 & 12 & 15 & 4/5 \\
S150V15-10 & 150 & 15 & 10 & 1.5 \\
S200V1-15 & 200 & 1 & 15 & 1/15 \\
S200V5-15 & 200 & 5 & 15 & 1/3 \\
\hline
\end{tabular}
\caption{Simulations performed for this paper.}
\label{simu-table}
\end{table}

\section{Results}
\label{res}
Our disc model being similar to the ``Steep'' disc used in \citetalias{2017MNRAS.467.1984G}, we use their simulation with $V_\mathrm{frag} = 15$~m\,s$^{-1}$  as a baseline to compare our results to.
Their simulation does not include any snow line, and a self-induced dust trap forms in a few hundred thousand years. A dust density enhancement starts to form at approximately 200~au (at 50,000~yr), drifts towards the star and stalls at $\sim 20$~au about 350,000~yr later. The resulting dust trap contains pebbles with a typical size of a few centimetres that are decoupled from the gas (see \citetalias{2017MNRAS.467.1984G} for a more detailed explanation of the evolution).

Our model with a snow line will impact the dynamics of dust grains through the three parameters we include: $V_\mathrm{fragin}$, $V_\mathrm{fragout}$ and $r_\mathrm{snow}$.
We fix the fragmentation thresholds and move the snow line in Section~\ref{snowpos}, then we study the fragmentation thresholds by focusing on the CO snow line in Section~\ref{CO}. Finally, we present the coupled effects of both the snow line location and the fragmentation thresholds in Section~\ref{groups}.

\begin{figure*}
\centering
\resizebox{\hsize}{!}{
\includegraphics{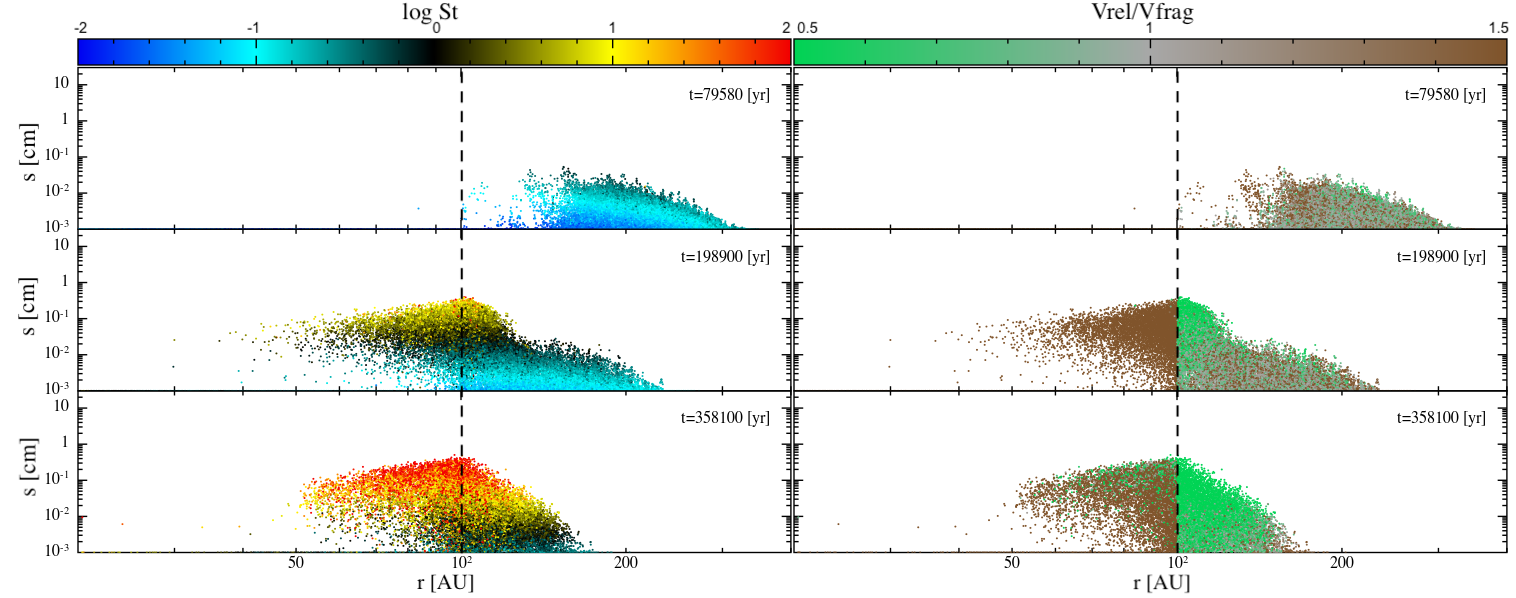}
}
\caption{Evolution of the dust size as a function of their radial distance to the star for simulation S100V1-15 for $\mathrm{t}=$~80,000, 200,000 and 360,000~yr. The left panel is coloured with the Stokes number and the right panel with the ratio $V_\mathrm{rel}/V_\mathrm{frag}$. The snow line is shown by the black dashed line on each panel.}
\label{evolution}
\end{figure*}

\begin{figure}
\centering
\resizebox{\hsize}{!}{
\includegraphics{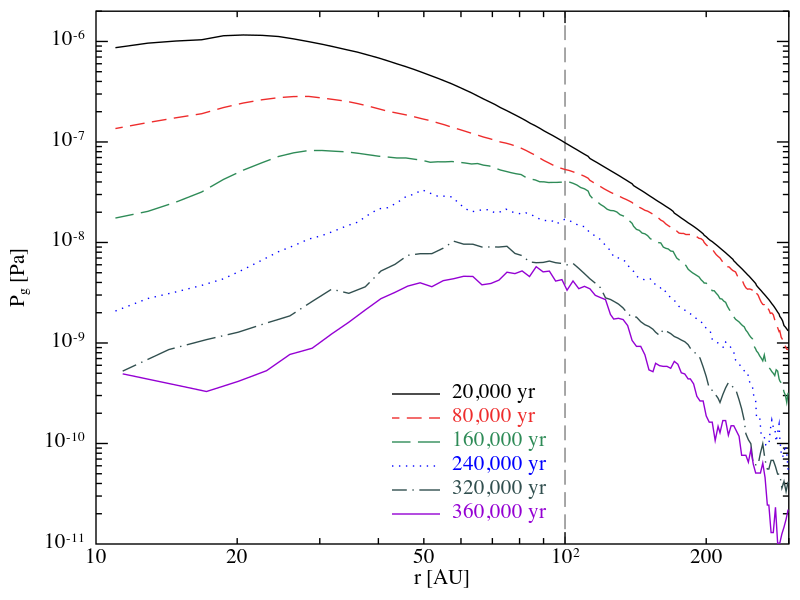}
}
\caption{Evolution of the gas pressure profile for simulation S100V1-15 at six different times between 20,000 and 360,000~yr.  The snow line is represented at 100~au by the grey dashed line.}
\label{evolution-Pg}
\end{figure}

First, we want to show the pile-up effect at the snow line mentioned in Section~\ref{intro}. In figure \ref{evolution} we show such an effect for simulation S100V1-15, where the difference in fragmentation velocities is large and the snow line is at what we define an ``intermediate'' distance from the star. For this simulation, the evolution is as follows. During the first 80,000 years, the grains in the outer region grow and drift inwards because $V_\mathrm{fragout}$ is large enough (top panel). Meanwhile, the grains interior to the snow line cannot grow due to the very small inner fragmentation velocity (1~m\,s$^{-1}$) and their larger relative velocities. When the outer grains reach sizes of a few mm ($\mathrm{St} \sim 1$), they cross the snow line at 100~au and enter a zone when the fragmentation threshold is 15 times lower. This low fragmentation velocity makes the grains fragment towards smaller sizes. In the process, their Stokes number can drop by an order of magnitude, which means that they drift slower (middle panel). As a result, the dust piles up at the snow line and enhances the local dust-to-gas ratio (bottom panel). Figure~\ref{evolution-Pg} shows the evolution of the pressure profile for the simulation. As the dust piles up near the snow line, the back-reaction onto the gas starts to pull it outside of the snow line (160,000~yr, green curve). A local pressure maximum, i.e.\ a self-induced dust trap, finally forms around 100~au, which will concentrate the grains at that location, lower their relative velocities below $V_\mathrm{fragin}$, and allow them to slowly grow without drifting. Note that the small wobbles at large distances from the star in the pressure profiles result from numerical noise.
This type of configuration (very large fragmentation velocity ratio and snow line at an intermediate distance) strongly affects the dust evolution because grains have intermediate sizes near the snow line and are marginally decoupled from the gas ($\mathrm{St} \sim 1$). As explained in Section~\ref{intro}, this means that they have the fastest radial drift and largest relative velocities, and so that they experience the fastest growth (if $V_\mathrm{rel} < V_\mathrm{frag}$) or fragmentation (if $V_\mathrm{rel} > V_\mathrm{frag}$).

Our aim is to understand under which conditions a self-induced dust trap forms around the snow line location, e.g for which values of $V_\mathrm{fragin}$, $V_\mathrm{fragout}$ and $r_\mathrm{snow}$. In the cases where that does not happen, we also examine the outcome of these other configurations.

\subsection{Effect of the snow line position}
\label{snowpos}
To understand the effect of the snow line location, we fix the discontinuity (e.g $V_\mathrm{fragin}$ and $V_\mathrm{fragout}$) and shift the snow line position. By doing this, we decouple the effect of the snow line position from the effect of the  {fragmentation thresholds}. In the following, we will refer to the change in fragmentation velocities as the ratio $V_\mathrm{fragin} / V_\mathrm{fragout}$. Even though this is useful to classify our simulations, we stress that it is degenerate and needs to be used with physical fragmentation velocity values (between a few and a few tens of m\,s$^{-1}$). We discuss in more details this choice of parameter in Section~\ref{groups}. In this Section, we take a fragmentation velocity ratio of 1/3 corresponding to $V_\mathrm{fragin} = 5$~m\,s$^{-1}$ and $V_\mathrm{fragout}=15$~m\,s$^{-1}$ respectively. We will refer to the simulations by their snow line position label only for ease of reading.
\begin{figure}
\centering
\resizebox{\hsize}{!}{
\includegraphics{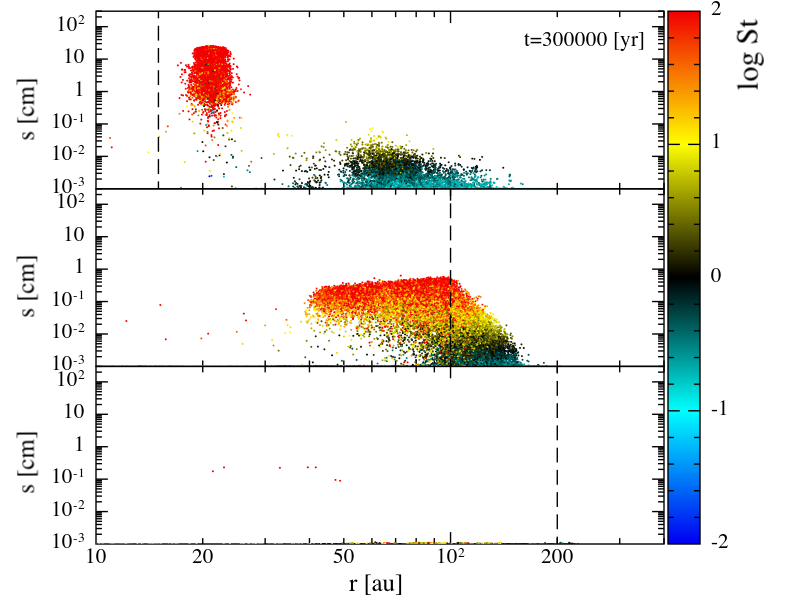}
}
\caption{Dust size as a function of their radial distance to the star after 300,000~yr for three simulations having the same discontinuity: S15V5-15, S100V5-15 and S200V5-15. The colour bar represents the Stokes number and the black dashed lines represent the snow line location for each simulations.}
\label{comp_pos}
\end{figure}

In figure \ref{comp_pos} we present the dust size distribution at 300,000~yr for 3 different simulations: S15, S100 and S200. Those simulations have three very different outcomes: a self-induced dust trap formed as if the snow line was not there (top panel, similar to Fig.~4 of \citetalias{2017MNRAS.467.1984G}); a dust trap that formed from the snow line location and that extends over $\sim 60$~au (middle panel); and no trap formation with only small grains due to a very efficient fragmentation in most of the disc (bottom panel).
These three behaviours correspond to three scenarios.
\begin{itemize}
\item[$\bullet$]{The snow line is close to the star (15~au, top panel): the dust trap forms exterior to the snow line at $\sim 20$~au as seen in \citetalias{2017MNRAS.467.1984G}. In this case, the snow line has no effect because the dust grains forming the trap never experience the zone where the fragmentation threshold is lower.}
\item[$\bullet$]{The snow line is at an intermediate distance (100~au, middle panel): the dust experiences the same type of evolution as simulation S100V1-15 (Fig~\ref{evolution}) and part of it is trapped at the snow line. Nonetheless, the discontinuity is less important than in simulation S100V1-15, which leads to a significant amount of the dust continuing to drift inward to radii of $\sim$ 40~au and slowly grow.}
\item[$\bullet$]{The snow line is far from the star (200~au, bottom panel): the dust reaches a low fragmentation velocity zone early in its evolution and is not spatially dense enough to grow efficiently and trigger the pile-up formation. Instead, the dust in the outer regions of the disc grows for a short time, drifts, and then fragments when it reaches the snow line without being able to grow after that.}
\end{itemize}

\begin{figure}
\centering
\resizebox{\hsize}{!}{
\includegraphics{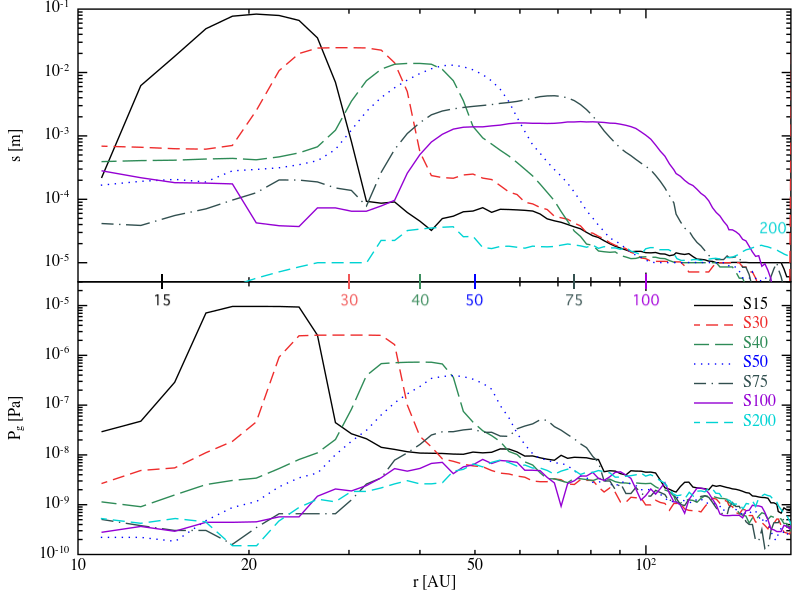}
}
\caption{Dust size (top) and gas pressure (bottom) radial profiles for simulations S20, S30, S40, S50, S75, S100 and S200 at 400,000~yr.}
\label{Pg_pos}
\end{figure}

The regimes where the dust is either trapped at the snow line or forms a structure extending from the snow line to the inner disc are of most interest, because it gives a direct link between two observables: the dust structures and the snow line. In Fig~\ref{Pg_pos} we compare 7 simulations with the same fragmentation velocity thresholds by showing their grain size and pressure profiles at 400,000~yr. Some of these simulations span a range of snow line positions (S30, S40, S50, S75 and S100) and result in the same category of dust distribution. The pressure maximum is closely following the snow line position for these simulations, as also indicated by the grain size profiles. The heights of these maxima are also correlated with the position of the snow lines. We observe a trend where the closer to the star the snow line is, the greater the pressure maximum. In fact, the pile-up starts to form in the outer disc and drifts towards the star, gathering dust in its path and enhancing its mass and density. Stopping the drift of grains with a snow line prevents the forming pile-up from collecting more dust, thus limiting its reservoir and by extension the intensity of the back-reaction onto the gas. For S100, there is a lack of correlation between the pressure maximum and the grain size distribution. In that case, the snow line shapes the dust into an extended dust ring containing similar sizes and spreading over several tens of au from $r_\mathrm{snow}$ inwards. This simulation, as opposed to the one we showed in Fig.~\ref{evolution}, indicates that the closer the fragmentation threshold ratio is to unity, the harder it is to trap dust at the snow line (see Section \ref{CO}). We also plotted S15 and S200, where no structures at the snow line are formed, for comparison purposes.

\begin{figure}
\centering
\resizebox{\hsize}{!}{
\includegraphics{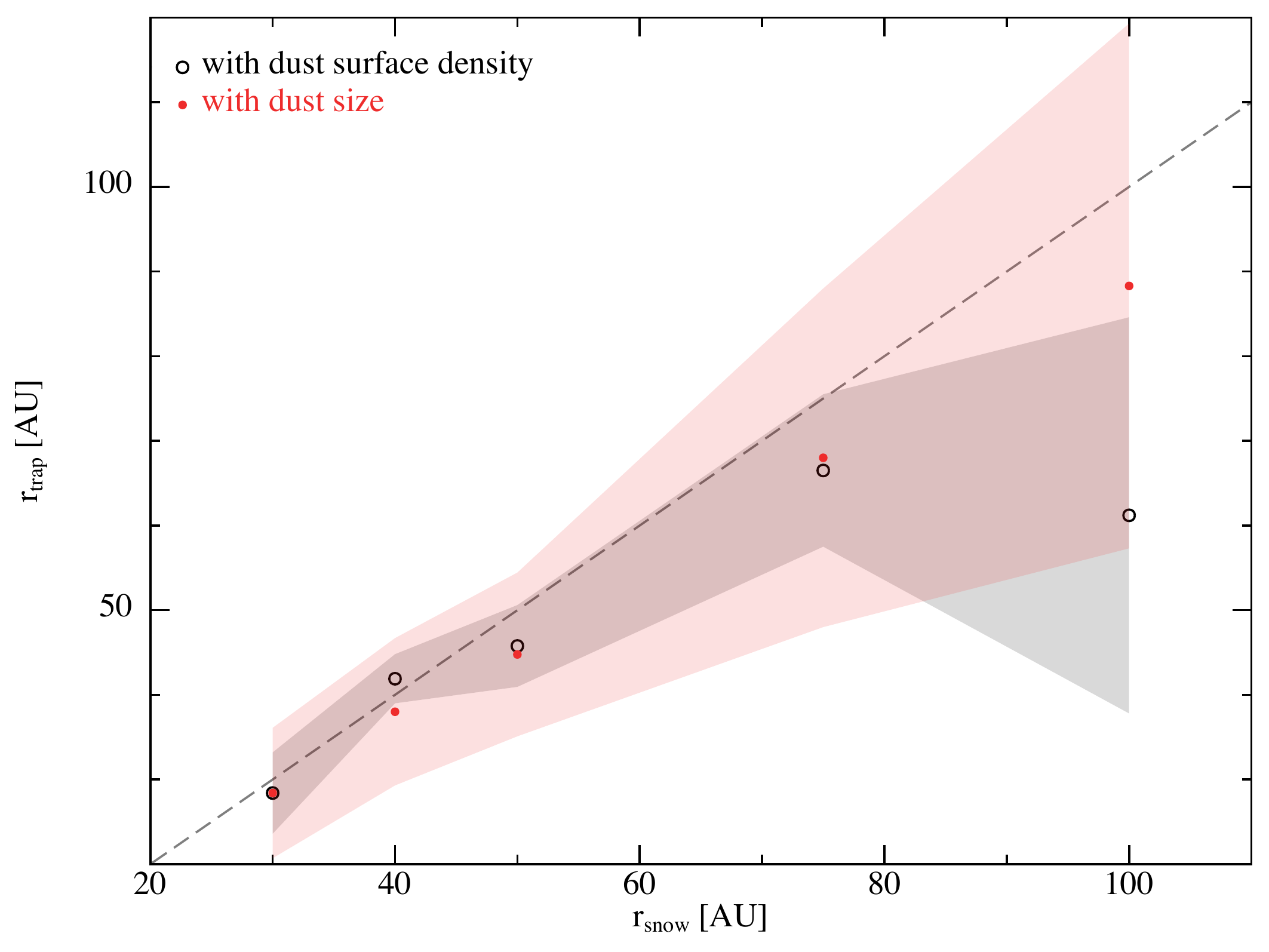}
}
\caption{Position of the dust trap as a function of the snow line location when estimated via the dust surface density maximum (black dots) and via the dust maximum size (red dots) after 350,000~yr for simulations S30, S40, S50, S75 and S100. The grey dashed line represents the case where the dust is perfectly trapped at the snow line ($r_\mathrm{trap} = r_\mathrm{snow}$). The shaded regions represent the width of the traps for each methods and are estimated where the given variable drops by 50\% on each side of the maximum.}
\label{width}
\end{figure}

Figure~\ref{width} demonstrates the ability for simulations S30, S40, S50, S75 and S100 to form dust traps that are rather well correlated with the snow line (shown by the grey dashed line and consistent with the width of these traps). We test this correlation with the grain size and dust surface density profiles of these simulations. We estimate the size of the uncertainty on both methods where the given variable drops by 50\% on each side of the maximum. For S30, S40, S50 and S75, whatever method that we use to identify the trap seems consistent. For S100, however, there is a discrepancy between the maximum dust surface density and the maximum grain size. While the maximum grain size is correlated with the snow line, the surface density reaches its maximal value around 60 au (black open circle at 100~au in Fig~\ref{width}). This indicates that dust trapping is harder the further away we are from the star, mainly because the amount of dust that can be gathered via inwards drift diminishes, which produces a less massive pile-up. The latter exerts a smaller back-reaction onto the gas and establishes a lower and wider gas pressure maximum. This is particularly the case when the snow line is located near the location where the gas pressure bump arises. Since the dust drift velocity is proportional to the gas pressure gradient (see Section~\ref{gd-dyn}), the timescale for dust accumulation in a trap is longer for a shallower pressure gradient. As a result, grains would be less efficiently trapped and be able to spread across larger distances interior to the snow line the further away $r_\mathrm{snow}$ is located from the star. For a given fragmentation velocity ratio, this means that the rings are wider at larger distances from the star, and narrower closer to it.

\subsection{Impact of the fragmentation thresholds: CO snow line}
\label{CO}

The position of the snow line is a key parameter in understanding how snow lines affect the dust dynamics. However, the way the grain composition impacts grain sticking properties also plays a major role in the evolution of self-induced dust traps around snow lines. To understand this, we focus this section on the physical CO snow line, which has been observed several times in recent papers (see references in Section~\ref{intro}). In our disc model, the CO snow line is located at $\sim 100 \mbox{ } \mathrm{au}$ for $T_\mathrm{subl} = 20 \mbox{ } \mathrm{K}$ \citep[][see equation~(\ref{rsnow})]{2013A&A...557A.132M}. We will keep this snow line position fixed to focus on $V_\mathrm{fragin}$ and $V_\mathrm{fragout}$. In this section, we will refer to the simulations by their fragmentation velocity couple for simplicity (e.g. ``1-15'' instead of ``S100V1-15'').

\begin{figure*}
\centering
\resizebox{\hsize}{!}{
\includegraphics{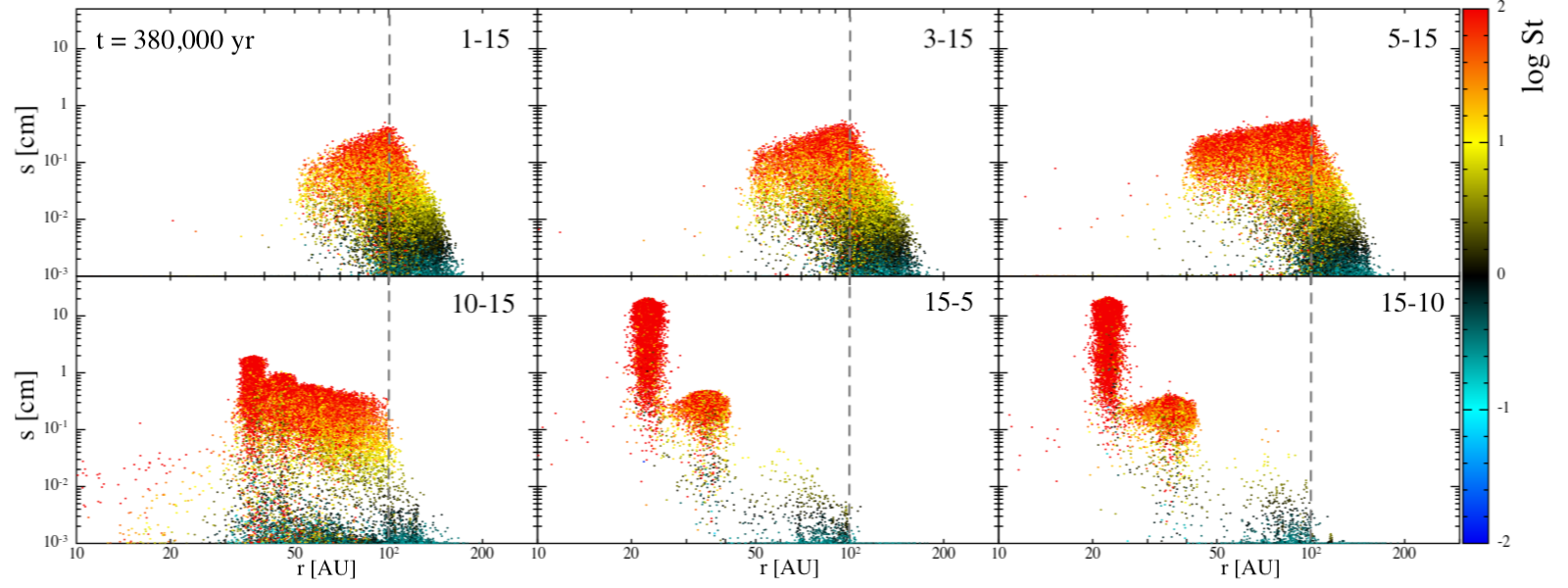}
}
\caption{Dust size as a function of radial distance to the star with $r_\mathrm{snow} = 100$~au (dashed grey line on each panel) at 380,000~yr, representing possible cases for the CO snow line. The label on the top right of each panel displays `$V_\mathrm{fragin}$-$V_\mathrm{fragout}$'. The colour scheme represents the Stokes number.}
\label{mosaique}
\end{figure*}

As the fragmentation behaviour of CO ice is uncertain, we tested six configurations, which we show in Fig.~\ref{mosaique}. We also examined the case where $V_\mathrm{fragin} > V_\mathrm{fragout}$ (15-5 and 15-10, last 2 panels), thus testing the possibility of CO aggregates diminishing the grains ability to stick as suggested by \citet{2017ApJ...845...68P}.

The first row of Fig.~\ref{mosaique} shows three simulations with a particularly low fragmentation velocity ratio ($V_\mathrm{fragin} / V_\mathrm{fragout} \leq 1/3$), which efficiently slows the dust drift at the snow line. For these simulations, the fragmentation velocity ratio is low enough to retain a significant amount of dust at $r_\mathrm{snow}$. However, we do see that when the fragmentation velocity ratio increases towards unity, the dust is more able to continue its drift while growing. The ensuing structures for these three simulations finally range from centred around the snow line (1-15) to extending from the snow line outside-in to $\sim 40$~au (3-15 and 5-15). The biggest grains are, however, always close to the snow line, showing that these simulations have a somewhat similar behaviour with different pile-up efficiencies at $r_\mathrm{snow}$.

\begin{figure}
\centering
\resizebox{\hsize}{!}{
\includegraphics{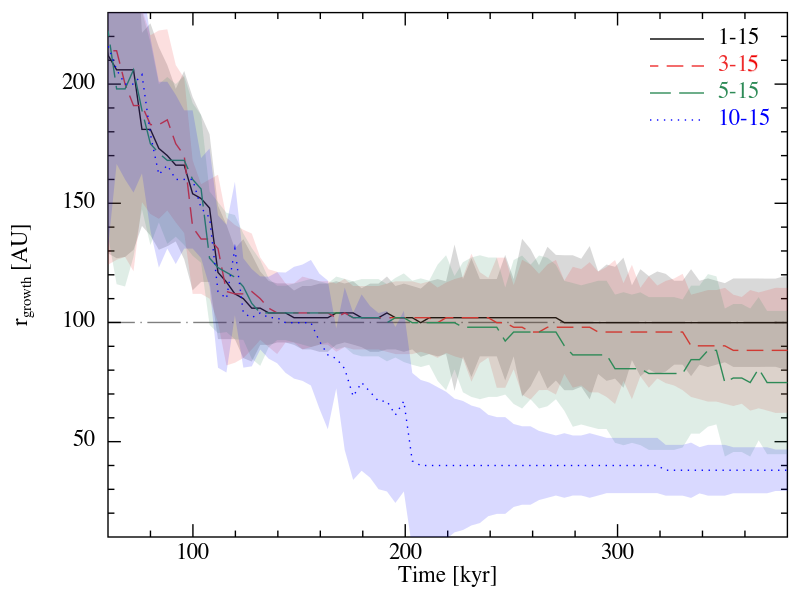}
}
\caption{Distance to the star of the maximum grain size (called $r_\mathrm{growth}$) as a function of time for the first four simulations presented in Fig. \ref{mosaique}. The shaded regions show the width $\Delta r_\mathrm{growth}$ of the maximum grain size profile, estimated where the size drops by 50\% on each side of the maximum. The snow line is represented by the grey dashed line.}
\label{SCOsA}
\end{figure}

\begin{figure}
\centering
\resizebox{\hsize}{!}{
\includegraphics{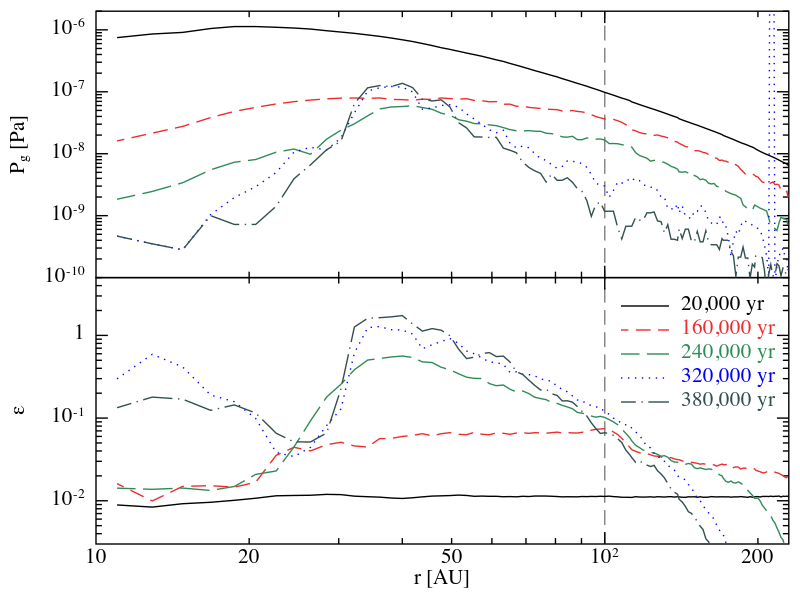}
}
\caption{Evolution of the gas pressure (top) and the vertically integrated dust-to-gas ratio $\varepsilon = \Sigma_\mathrm{d}/\Sigma_\mathrm{g}$ (bottom) profiles at 30,000, 160,000, 240,000, 320,000 and 380,000~yr for simulation S100V10-15. The snow line is represented by the grey dashed line.}
\label{evol-profiles}
\end{figure}

In Fig. \ref{SCOsA} we show the evolution of the position of the largest grains (measured with the dust size profile) as well as the associated width as a function of time for simulations 1-15, 3-15, 5-15 and 10-15 (essentially with a fragmentation velocity ratio inferior to 1). We clearly see the slight deviation from the snow line position when the inner fragmentation velocity increases (from 1-15 to 5-15). However, in these 3 simulations the dust distribution seems well correlated with the snow line.
For 10-15 (fourth panel of Fig. \ref{mosaique} and blue curve of Fig. \ref{SCOsA}), on the other hand, the fragmentation velocity ratio is not sufficiently low to trap dust at the snow line. However, it slows grain growth and creates a similar extended dust structure as simulation 5-15, with the major difference that this time the biggest grains are not at the snow line. With that in mind, there appears to be a shift in the dust behaviour between the third and fourth panels of Fig. \ref{mosaique} (for this particular value of $r_\mathrm{snow}$). The 10-15 simulation (fourth panel) is presented in Fig.~\ref{evol-profiles}. One sees that a small enhancement in the dust-to-gas ratio shifts from the outer disc towards the star because of the grains drifting from the outer disc as they grow (the dust evolution in the outer disc is the same as that shown in Fig~\ref{evolution} for simulation S100V1-15). We first see a `bump' in the dust-to-gas ratio at 160,000~yr (red dashed line) at 100~au, corresponding to the dust reaching the snow line and being slightly slowed in their growth. Indeed, as the fragmentation velocity ratio approaches unity (the fragmentation velocities are only separated by $5$~m\,s$^{-1}$ ), grains are able to continue their drift and growth but more slowly, until they are finally decoupled from the gas with a peak between $30$ and $40$~au after 320,000~yr (blue curve) and extending out to the snow line. At this point, one sees a high pressure maximum and a dust-to-gas ratio of order unity in the inner part of the trap, decreasing with the distance to the star until right outside of the snow line. Additionally, the bottom panel of Fig~\ref{evol-profiles} shows the distance to the star where the dust-to-gas ratio drops off. This highlights the dust disc shrinking over time from 300~au at the moment of the dust injection to $\sim 170$~au at 380,000~yr.

\begin{figure}
\centering
\resizebox{\hsize}{!}{
\includegraphics{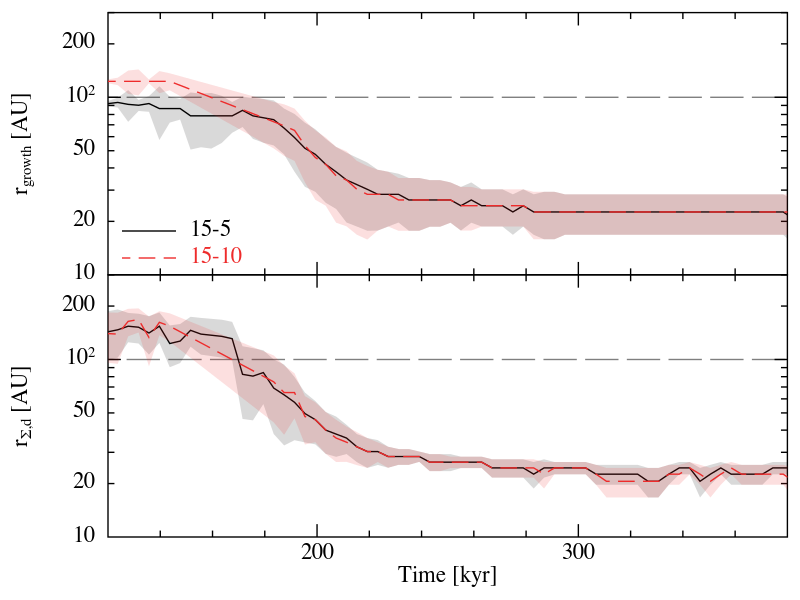}
}
\caption{Evolution of the locations of the maximum dust size (top) and maximum dust surface density (bottom) as a function of time for simulations 15-10 and 15-5 (last two panels of Fig. \ref{mosaique}). The shaded regions show the width of these profiles and are estimated where there is a 50\% decrease each side of the maxima. The snow line is represented by the grey dashed line on each panel.}
\label{SCOb}
\end{figure}

\begin{figure}
\centering
\resizebox{\hsize}{!}{
\includegraphics{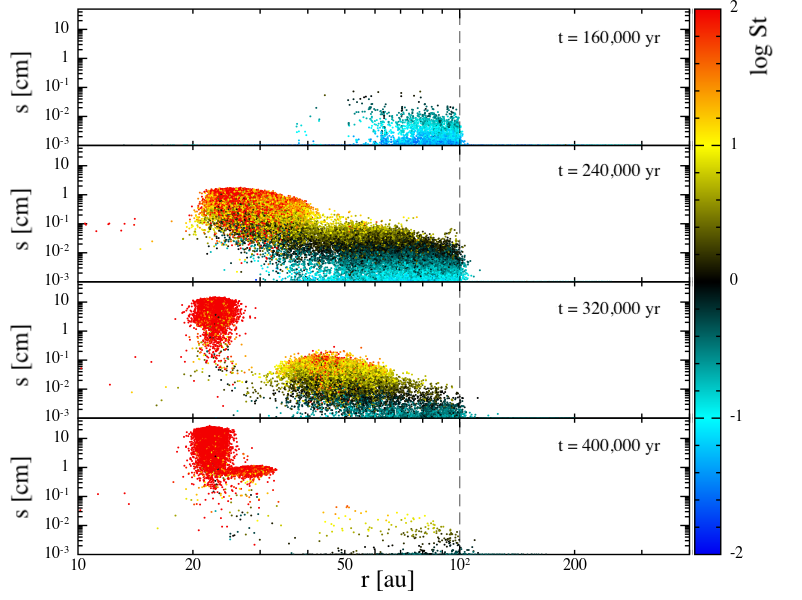}
}
\caption{Dust size as a function of their radial distance to the star at four different times for simulation 15-5. The colour bar represents the Stokes number and the grey dashed line represents the snow line location.}
\label{15-5-evol}
\end{figure}

On the contrary, when $V_\mathrm{fragin} / V_\mathrm{fragout} > 1$ (last two panels of Fig~\ref{mosaique}), the dust behaviour is quite different. The lower value of the outer fragmentation threshold keeps dust at smaller sizes but lets it slowly drift towards the star and settle to the mid-plane. When grains finally cross the snow line, the over-density (bottom panel of Fig.~\ref{SCOb}) created by the settling and the slow radial drift contains dust particles that become free to grow as the fragmentation threshold becomes much larger (top panel). In these simulations, dust growth starts either just outside (15-10, red curve) or inside (15-5, black curve) the snow line and the dust drifts towards the star to form a trap at $\sim20$~au, containing cm-sized grains. These self-induced dust traps are similar to those found in \citetalias{2017MNRAS.467.1984G} without a snow line, although they are between 10 to 15\% less massive due to the fact that they gather less material along the way.
Figure~\ref{15-5-evol}, displaying the evolution of the radial grain size distribution for simulation 15-10, also shows that while the final stages show no relation to the snow line, earlier stages clearly do, in the form of a sharp cut-off.

\section{Discussion}
\label{discu}

\subsection{Self-induced dust traps and snow lines: a parameter study}
\label{groups}

\begin{figure}
\centering
\resizebox{\hsize}{!}{
\includegraphics{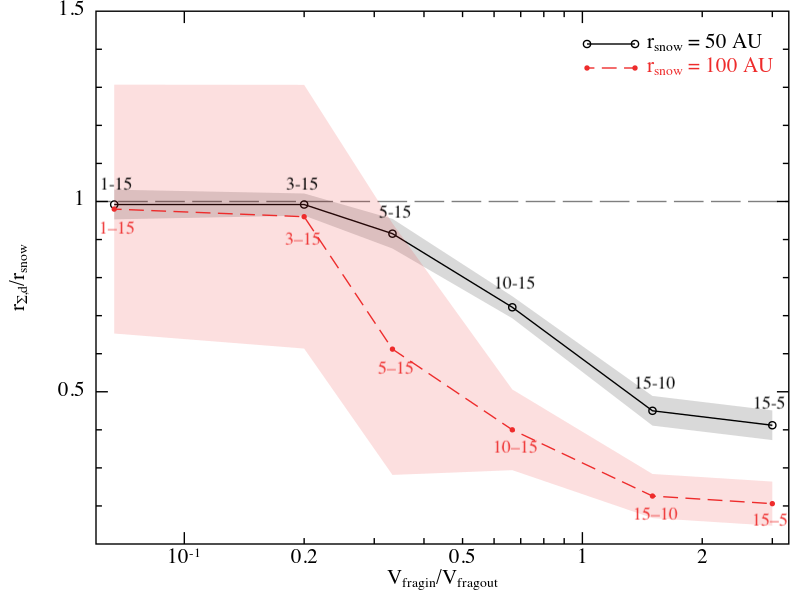}
}
\caption{Position of the maximum dust surface density normalised to the snow line position as a function of the fragmentation velocity ratio for two sets of snow line positions. The shaded regions are showing the width of the dust surface density profile around the maximum and are estimated where the maximum drops by 50\% on each side of the maximum. The label next to each dot represents ``$V_\mathrm{fragin}$-$V_\mathrm{fragout}$'' for the corresponding simulation. The snow line is represented by the grey dashed line.}
\label{rsnow-vfrag}
\end{figure}

We investigated the effects of the fragmentation velocities and the snow line location separately in Sections~\ref{CO} and \ref{snowpos} respectively, which leads us to consider the interplay between the two. In Fig.~\ref{rsnow-vfrag}, we can see a detailed answer to this question for two snow line locations (50~au and 100~au respectively). Even though these models behave similarly (i.e. they both trap dust at their respective snow line), they have slight differences, which makes them interesting to compare.
The radial position of the maximum dust surface density as well as its width gives us an indication of the trapping efficiency of the snow line with respect to the fragmentation velocity ratio. For a very low ratio (1-15 and 3-15), the dust is strongly piled-up at the snow line, independent of its position. For a fragmentation velocity ratio of 1/3 (5-15), the dust starts to drift towards the star for both snow lines. However, since the largest grains are still at the snow line and the width of the dust maximum surface density coincide with the snow line, we still associate these simulations with a pile-up at $r_\mathrm{snow}$.

For larger ratios (10-15), as we discussed in Section \ref{caseii}, the pile-up at the snow line does not occur and the dust slowly drifts inwards of $r_\mathrm{snow}$ for both snow line positions. For $r_\mathrm{snow} = 100$~au, the resulting rings are significantly narrower than cases 1-15, 3-15 and 5-15.
For even larger ratios (i.e. $V_\mathrm{fragin} > V_\mathrm{fragout}$), dust can start growing at the snow line and evolve towards the inner region of the disc. This happens for both snow lines and never results in a pile-up at its location.

\begin{figure}
\centering
\resizebox{\hsize}{!}{
\includegraphics{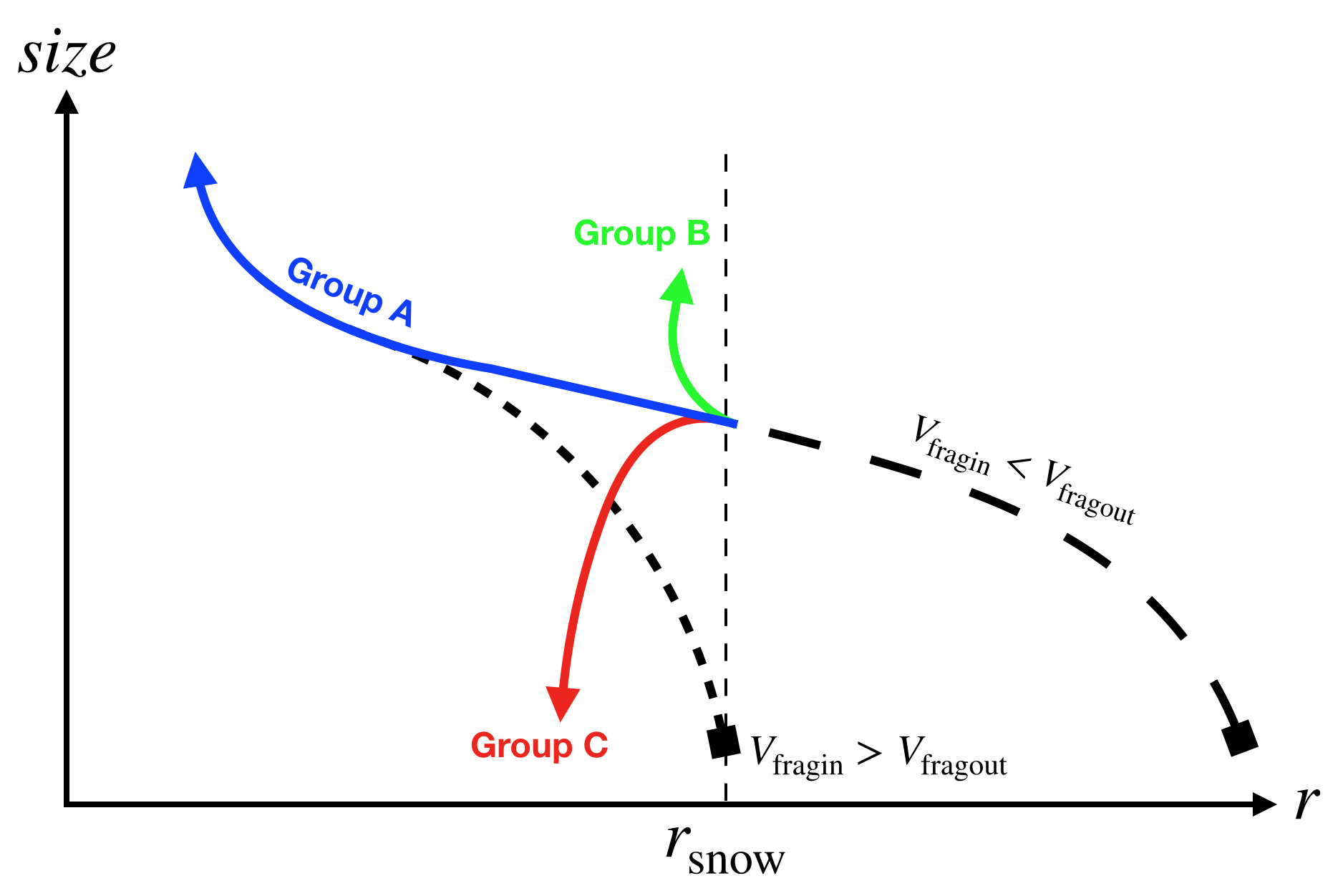}
}
\caption{Evolution of a dust trap in the ($r$,$s$) plane depending on the parameters in the simulation. Group A (blue) leads to ``usual'' self-induced dust traps that end up at a position separate from the snow line. Group B (green) leads to an efficient dust trapping at the snow line location. Group C (red) leads to the trap's self-destruction. The thick black dashed lines represent the initial growth history of dust if $V_\mathrm{fragin} > V_\mathrm{fragout}$ (short dashed) or the opposite (long dashed). The thin black dashed line represents the snow line.}
\label{analyse2}
\end{figure}

In these 12 simulations, for a given fragmentation velocity ratio, the effect seems to be generally the same for different snow line positions. Only the growth timescale and eventually the width of the traps are changed with the distance to the star. More generally, with all the simulations listed in table \ref{simu-table}, we can identify three different behaviours regarding dust growth and trapping around a generic snow line. We gathered these behaviours into groups that correspond to different types of outcomes in our simulations. The typical history of a forming dust trap with respect to these groups is represented in figure \ref{analyse2}, which is meant to summarise our analysis. Group A (blue) does not create a dust trap at the snow line location but rather in the inner disc where the dust would slow its drift after decoupling from the gas. Group B (green) is the most interesting group, because a self-induced dust trap forms at the snow line location and allows the dust to slowly grow without drifting. Group C (red) is the most affected group, since the snow line inhibits the dust trap formation. The detailed mechanism for each group is as follows:

\begin{itemize}
\item{Group A: The dust is not trapped at the snow line and continues its course towards the star to form a self-induced dust trap between 20 and 50~au. This is because the discontinuity is close to 1 or because $V_\mathrm{fragin} > V_\mathrm{fragout}$. Hence, either the dust is able to grow interior to the snow line, or the small change in fragmentation velocity is insufficient to stop its radial drift. This group creates self-induced dust traps that are the closest to those in \citetalias{2017MNRAS.467.1984G}.}
\item{Group B: $V_\mathrm{fragout}$ is large enough compared to $V_\mathrm{fragin}$ and the snow line is approximately between 30 and 130~au from the star. It can thus efficiently stop the drift of grains at $r_\mathrm{snow}$ that have had sufficient time to grow from the outer disk and start to decouple from the gas, as explained in Fig~\ref{evolution}. When they enter the region interior to the snow line, the Stokes number of the largest grains exceeds unity (middle-left panel of Fig~\ref{evolution}), they slow down their drift and pile up. This reduces their relative velocity (bottom-right panel of Fig~\ref{evolution}), which prevents them from fragmenting even with the lowest fragmentation threshold. In this group, the trap forms at the snow line and the grains reach sizes of several mm in 400,000~yr. We call them self-induced dust traps, but we emphasise that they contain less material due to the fact that the pile-up gather less mass along its smaller course. The dust contained in these kind of traps are growing slower than those in Group A due to their larger distance to the star (see Appendix~\ref{tgrowthapp}).}
\item{Group C: $V_\mathrm{fragout}$ is large enough compared to $V_\mathrm{fragin}$, the snow line is at approximately 130~au or beyond and $V_\mathrm{fragin}$ is low enough so that grain experience fragmentation when entering the inner zone. As the typical grain growth timescale $\tau_g$ increases as a function of the radius (see Appendix~\ref{tgrowthapp}), grain growth is stopped before reaching sizes corresponding to $\mathrm{St}\sim 1$. As a result, grains cannot pile-up at the snow line, meaning that the dust is forced to fragment towards smaller sizes and the dust-to-gas ratio is not sufficiently large to allow the back-reaction to have an important effect.}
\end{itemize}

\begin{figure}
\centering
\resizebox{\hsize}{!}{
\includegraphics{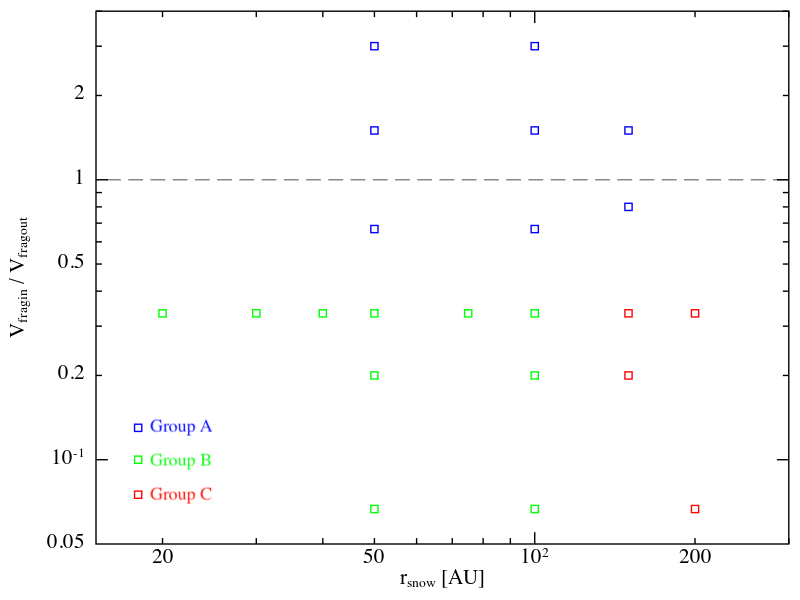}
}
\caption{Positions of the simulations performed for this paper in the  ($r_\mathrm{snow}$,$V_\mathrm{fragin}/V_\mathrm{fragout}$) plane. The simulations split into three groups that are detailed in Section~\ref{groups}. The grey dashed line represents the limit where $V_\mathrm{fragin} = V_\mathrm{fragout}$.}
\label{analyse}
\end{figure}

In figure \ref{analyse}, we show all simulations performed for this paper and sort them in these three groups. The y-axis represents the fragmentation velocity ratio $V_\mathrm{fragin} / V_\mathrm{fragout}$, while the x-axis represents the snow line location. One clearly sees that these groups occupy specific regions in this plane, given the previous explanations.
It appears the snow line position only has a `trigger' effect, where above a certain distance to the star (between 100 and 150~au) the trap formation is shut off for fragmentation velocity ratios able to form traps closer in.

In our snow line model, we found that for a significant range of fragmentation velocities and snow line positions, there is an efficient pile-up of dust at the snow line location which will lead to a dust trap. Despite this efficiency, the trap formation depends on the values of the fragmentation velocities, which are still debated. For this particular disc model, we found that dust is able to grow in some parts of the disc as long as the fragmentation velocity (in either one of the zones) is of the order of 10 - 15 $\mathrm{m.s^{-1}}$.

Our analysis of the role of $V_\mathrm{fragin}$, $V_\mathrm{fragout}$ and $r_\mathrm{snow}$ distinctively highlights the interplay between the formation of self-induced dust traps and the presence of a snow line. Depending on the snow line characteristics, it could either be a favorable location and help dust trapping or it could interfere with trap formation. For group C, the snow line needs both to be far from the star and to correspond to a large difference in the grain sticking properties, which is something that we don't see nor we expect to happen in discs. The use of the fragmentation velocity ratio $V_\mathrm{fragin} / V_\mathrm{fragout}$ as a discriminant is subject to discussion, since it is degenerate. However, with probable values for the fragmentation velocities, we find that it is a rather good indicator of the dust fate in our simulations. Moreover, this problem has many degrees of freedom, which means that every parameter we could choose would be degenerate as well. We made the choice of using this one to facilitate interpreting our results. We stress that the specific values for the fragmentation velocities and the snow line positions apply for this particular disc model and are not universal, even though the steep disc profile is an `average' disc model from observations \citep[see, e.g.,][]{2014ApJ...788...59W}. Overall, this analysis is useful to point out tendencies in the behaviour of dust growth alongside snow lines in protoplanetary discs. We expect other disc models to behave in a similar way.

The self-induced dust trap mechanism has been seen consistently with our code and for different disc models \citepalias{2017MNRAS.467.1984G}. Very recently, \citet{2019arXiv190607708G} found that the dust could revert the gas flow because of its back-reaction on the gas, which is similar to the self-induced dust traps. However, they pointed out that they could not find the natural pile-up mechanism due to the dust growing and decoupling from the gas. This discrepancy might be due to two major differences between our two groups.
\begin{itemize}
\item{The numerical methods: our simulations are 3D, Lagrangian, self-consistently compute the forces on gas and dust SPH particles, and numerically integrate the equations of motion, as opposed to the Eulerian, grid-based methods used by many authors, which rely on semi-analytical models for the evolution of both gas and dust.}
\item{The growth and fragmentation models: we use a locally monodisperse approach for each SPH particle producing a size distribution in small volumes (see Figs.~\ref{evolution}, \ref{comp_pos}, \ref{mosaique} and \ref{15-5-evol}), while \citet{2019arXiv190607708G} and other authors solve the Smoluchowski equation for multiple dust sizes as in \citet{2010A&A...513A..79B} or use the simpler two-population model of \citet{2012A&A...539A.148B} based on the former. In particular, the fragmentation is stronger in our model than in theirs, which produces a top-heavy size distribution and is more similar to erosion. This might be important for the self-induced dust trap mechanism, because having a steep gradient of grain sizes leads to a steep gradient of radial drift velocity, thus reinforcing a potential pile-up.}
\end{itemize}
A complete comparison between grid-based and SPH methods has been started and will certainly give us more answers regarding this discrepancy.

\subsection{To be or not to be at the CO snow line ?}  

While the water snow line is thought to be linked to dust structures, figuring out if the same is true for CO is one of our objectives. The recent observations of the CO snow lines in HD~163296 \citep{2013A&A...557A.132M} and HD~169142 \citep{2017ApJ...838...97M} provide some clues, but more are needed. With our simulations, we find that the dust evolution is highly dependent on the difference in grain composition, i.e.\ in fragmentation thresholds across the snow line. As we do not know for sure how CO affects the dust mantle, we consider 3 cases.

\subsubsection{Case i: $V_\mathrm{fragin} \ll V_\mathrm{fragout}$}

In this case, the inner grains fragment much more easily than the outer ones. This means that when CO freezes-out on the surface of grains, it strongly increases the energy necessary to break their mantle. As a result, the snow line has the effect of trapping the grains as seen in Section~\ref{CO}. Here, one should see correlations between a dust surface density maximum and the CO snow line. \citet{2018ApJ...869L..42H} computed the mid-plane temperature as a function of the distance to the star for 18 discs in the DSHARP project using a passively irradiated disc model and linked the dust surface density to different snow lines (including CO). In these 18 discs, 5 of them seem to have a `bump' near the CO snow line (namely HD 163296, Elias 24, HD 143006, Elias 20 and RU Lup). Such bumps are similar to the ones in simulations 1-15, 3-15 and 5-15 (first row of Fig. \ref{mosaique}).
However, it seems unlikely that CO would change the dust behaviour so dramatically in protoplanetary discs, especially considering the fact that we expect CO to be mixed with water ice, which is thought to have a high fragmentation velocity. Additionally, such a strong change would be seen in every disc, which is not the case in the DSHARP sample or for other ALMA observations. We would, as a result, argue that the observed bumps are probably not due to a strong change in the dust sticking properties across the CO snow line. 

\subsubsection{Case ii: $V_\mathrm{fragin} \lesssim V_\mathrm{fragout}$}
\label{caseii}

If the fragmentation velocity ratio is closer to unity (e.g. $2/3$, 4th panel of Fig.~\ref{mosaique}), the dust is not located at the snow line, but extends radially from a few tens of au out to the snow line. In this scenario, CO affects the dust behaviour by slightly increasing its ability to grow when frozen-out. As seen in our simulations, this slight change can significantly impact the dust evolution. In this case, the correlation between the dust and the snow line is less obvious. While there is no visible dust trapping at the snow line, the dust surface density decreases in the outer disc. Since the dust is slowed in its growth and drift by the snow line but is not trapped, the dust eventually extends from a few tens of au out to the snow line. As a result, it produces a drop-off of the dust surface density profile just outside the snow line, without any visible pile-up near the condensation front. In the DSHARP data \citep{2018ApJ...869L..42H}, a lack of dust pile-up at the snow line with a density profile dropping-off after the snow line seems probable for 6 discs (namely WaOph 6, MY Lup, WSB 52, Sz 114, Sz 129 and GW Lup) and is somewhat similar to our simulation 10-15.

It is worth noting that we did not take into account the diffusion of sublimated CO inwards and outwards of the snow line. However, it has been proven (for water) that the snow line can lead to a diffusion of material that could enhance the dust surface density just outside the snow line \citep{2017A&A...608A..92D}. In their paper, they used a model similar to ours for the fragmentation thresholds and added the diffusion terms. They found that the fragmentation thresholds difference dominated the dust behaviour in discs at the water snow line, which is likely the snow line with the largest difference between $V_\mathrm{fragin}$ and $V_\mathrm{fragout}$ (with 1~m\,s$^{-1}$ for bare silicates and 10 to 15~m\,s$^{-1}$ for icy aggregates). Even though CO is less abundant than water, the diffusion could decrease the amount of solid material just inside the snow line at the benefit of increasing the CO dust surface density just outside of it \citep{2017A&A...600A.140S}. In that paper, the authors also found that the dust size was not significantly increased just outside of the CO snow line, but that the abundance of CO itself could be enhanced by a factor of a few just inside of the ice line. This would mean for our study that a fragmentation velocity ratio closer to unity could still be efficient enough to capture dust at the snow line. In that sense, our simulations may be perceived as somewhat pessimistic for dust trapping around the CO snow line.\\

\subsubsection{Case iii: $V_\mathrm{fragin} > V_\mathrm{fragout}$}

In this case, the inner grains are more resistant to fragmentation than the outer ones. This means that when CO freezes-out on the surface of grains, it weakens them. This has been proposed by \citet{2017ApJ...845...68P}, where they chose to assimilate the behaviour of CO$_2$-covered grains with that of silicates, i.e. that their fragmentation velocity is of the order of 1~m\,s$^{-1}$. Here, we consider a similar behaviour for CO.
However, at these distances from the star, the grains are not only covered with CO, but rather a mixture of $\mathrm{H_2O}$, CO, $\mathrm{CO}_2$ or even $\mathrm{NH}_3$. This indicates that while CO would diminish the mantle's ability to stick, it would still be relatively high due to the other elements (mainly the water ice which sticks efficiently). The two cases we tested are 15-5 and 15-10 and in these simulations, the dust is unable to start growing in the outer disc but rather starts growing near the snow line.
As a consequence, the growing dust never piles up at the snow line but drifts from it towards the inner parts of the disc. This means that the large grains and most of the dust mass drift inside of the snow line, which translates into a sharp cut-off for the surface density profile interior to the snow line. We also should not find large grains exterior to the CO snow line because grains are not allowed to grow there. This is consistent with what \citet{2017ApJ...845...68P} found in their model II for $\alpha=10^{-2}$ (the closest to our model), where dust growth only happens between the water ice line and the $\mathrm{CO_2}$ line. However, as they do not take back-reaction into account on the gas evolution, they do not see any decoupling of the dust with respect to the gas, as opposed to us. As a result, they see dust extending from the $\mathrm{CO_2}$ ice line to the water ice line. In our case, the results of this group of simulations largely differ from the 10-15 case where the dust starts its growth from the outer disc (exterior to the snow line). In the DSHARP project, 3 discs seem compatible with this situation (namely HD~142666, DoAr 33 and SR 4), and a lack of large grains exterior to the CO snow line is also consistent with what \citet{2016A&A...588A.112G} and \citet{2019ApJ...881..159M} observed in HD~163296 and HD~169142, respectively.

\subsubsection{From simulations to observations}
\label{sim-obs}

At the end of our simulations, small grains are depleted at the benefit of large grains, which differs from most observations, from which we would expect the survival of a population of small grains throughout the disc. This is a consequence of our growth and fragmentation model, which considers the size distribution to be highly peaked locally. As a result, this model favours large grains at the expense of small ones, which is appropriate when investigating dust trapping at large Stokes numbers. Computing dust growth and fragmentation with the full size spectrum would require the resolution of the Smoluchowski equation \citep{1916ZPhy...17..557S}. This is incredibly challenging within the SPH formalism, but is in progress (M. Lombart, private communication). The collisions between non equal mass particles would help replenishing smaller sizes and thus produce a smoother dust surface density profile, which would be more comparable to observations. However, our model is useful in order to track the maximum of the dust bulk mass, which becomes trapped.

Among our 3 cases, 14 out of 18 discs in the DSHARP program seem to carry similar signatures to what we would expect. However, we can not directly compare our simulations with observation, as this requires radiative transfer, which we will explore in forthcoming work.
Nevertheless, we can discuss our preliminary findings. Case i seems unlikely for the CO snow line, because it would require a large difference in the grains sticking properties. We expect this difference to be more subtle for carbon monoxide. This is, however, what we would expect for the water snow line. This reduces the number of discs carrying similar signatures to our study to 9 out of 18. As \citet{2018ApJ...869L..42H} pointed out, it is difficult to draw any satisfying conclusion from a signature only seen by a subset of these discs. However, it is crucial to know if dust growth starts outside, at or inside the snow line.

From our simulations, we see that the dust behaviour is largely dependent on the difference in sticking properties on either side of the snow line (e.g. a change of 5~m\,s$^{-1}$  can result in vastly different dust distributions). 
Additionally, the physical state of CO depends on the temperature, pressure and chemical structure of the disc and the CO molecule has rather complex distributions as it goes through multiple chemical reactions \citep{2014ApJ...783L..28M,2016ApJ...816L..21C}. The conditions may vary from one disc to another because of the different stellar host or of the composition of the molecular cloud it originated from. It is certainly possible that the local abundance of CO plays a part in the fragmentation velocity threshold in our model. We could imagine that different CO distributions would lead to the grains behaving differently when they cross the snow line. This could explain why we cannot find a consensus amongst all the observed discs: they would not share identical CO distributions. A full chemical and dynamical study is required to explore that idea in more detail. More generally, making detailed comparisons with observations requires dedicated simulations for each disc, taking into account its particular structure and location of the CO snow line.

\subsection{Planet(esimal) formation}
\label{planetesimal}
Concentrating dust in traps is what planet formation theories need to save dust from being accreted onto the star. At the end of our simulations, the biggest grains have typical sizes of 1~m. To continue dust growth to larger sizes, we would need to take self-gravity into account. With such pile-ups, the streaming instability \citep{2005ApJ...620..459Y} can transform pebbles into planetesimals as long as the disc is not too viscous. For a full comparison between self-induced dust traps and the streaming instability, we refer the reader to the discussion in \citetalias{2017MNRAS.467.1984G}.
Self-induced dust traps are a natural way of trapping dust in rings, where the dust-to-gas ratio is larger than the classic value of 1\% by one or two orders of magnitude. With such enhancements of the dust density compared to the gas, it is possible that the streaming instability and self-induced dust traps could be working together to form planetesimals in pressure bumps \citep{2018MNRAS.473..796A}. In particular, these authors found that the streaming instability can develop in a pressure bump for discs with a higher viscosity than previously thought ($\alpha \gtrapprox 10^{-3}$) at the cost of a slower growth rate. This is encouraging for the early stages of planet formation. 

While our grains are considered compact, it has also been shown that porosity \citep{2013A&A...557L...4K} can act in favour of planet formation \citep{2012ApJ...752..106O}. The porosity of grains increase their collisional cross section and can lead to a faster growth rate and a slower radial drift when they enter the Stokes drag regime \citep[][Garcia \& Gonzalez, submitted]{garcia:tel-01977317}. A more complete model of dust evolution with grain growth, fragmentation, porosity and snow lines would be the next step.\\

\section{Conclusion}
\label{conclu}

Self-induced dust traps are the result of a large number of dust particles growing and piling up because of their collective effect on the gas.
We showed that snow lines affect the dust dynamics through dust growth and fragmentation and can lead to an efficient self-induced dust trapping at a specific location.
We summarise our main findings as follows:
\begin{itemize}
\item{The self-induced dust trap mechanism is robust: it happens with sharp differences in the fragmentation velocity at various locations. It forms cm to m sized grains which are decoupled from the gas and are safe from both the radial drift and the fragmentation barriers. They are distributed in radial concentrations with dust-to-gas ratios close to unity.} 
\item{A rather small fragmentation velocity difference (typically $5$ $\mathrm{m.s^{-1}}$) can result in vastly different dust distributions. The snow line pile-up efficiency seems strongly dependent on the grain surface composition (which is represented by the fragmentation velocities).}
\item{ALMA images of discs at millimetre wavelengths \citep[e.g.][]{2016A&A...588A.112G,2018ApJ...869L..41A} show similar features to our simulations and may suggest that these discs have different CO structures.}
\item{Even when there is no link between the dust structures and the CO snow line at later stages of evolution, our simulations show that dust growth could have started near the snow line at earlier stages.}
\item{More generally, the further the snow line is from the star, the more it hinders dust growth up to a point where growth is no longer possible ($\sim 130$~au for $V_\mathrm{frag,max} = 10$~m\,s$^{-1}$  in our disc).}
\item{The weaker the inner grains are compared to the outer ones with respect to fragmentation, the more efficiently dust piles up at the snow line.}
\end{itemize}

Taking into account the effects of snow lines on dust growth is a step towards a better understanding of planet formation.
Our next step will be to process our simulations with a radiative transfer code such as \textsc{mcfost} \citep{2006A&A...459..797P}. By doing so, we will translate our simulations into observational signatures around snow lines and confirm the likelihood of our results compared to previous ALMA disc images.

\section*{Acknowledgements}
AV would like to thank Sarah T. Maddison and the anonymous referee for {their} help in the improvement of this paper. Anthony J.L. Garcia is also to be thanked for meaningful discussions. This research was supported by the \'Ecole Doctorale PHAST (ED 52) of the Universit\'e de Lyon. The authors acknowledge funding from ANR (Agence Nationale de la Recherche) of France under contract number ANR-16-CE31-0013 (Planet-Forming-Disks) and thank the LABEX Lyon Institute of Origins (ANR-10-LABX-0066) of the Universit\'e de Lyon for its financial support within the programme `Investissements d'Avenir' (ANR-11-IDEX-0007) of the French government operated by the ANR. This project has received funding from the European Union's Horizon 2020 research and innovation programme under the Marie Sk\l{}odowska-Curie grant agreement No 823823. SPH simulations were run at the Common Computing Facility (CCF) of LABEX LIO. All figures except Figs~\ref{snowlinescheme} and \ref{analyse2} were made with SPLASH \citep{2007PASA...24..159P,2011ascl.soft03004P}.



\bibliographystyle{mnras}
\bibliography{refs}




\appendix
\section{Growth time scale throughout the disc}
\label{tgrowthapp}

The typical growth time scale, $\tau_g$ can be estimated by writing:
\begin{equation}
\tau_g = \dfrac{s}{\dfrac{\dd s}{\dd t}}  = \dfrac{s}{\dfrac{\rho_\dd}{\rho_{\mathrm{s}}}V_\mathrm{rel}}.
\end{equation}
With the power law formulation of the Stokes number:
\begin{equation}
\mathrm{St} = \dfrac{\rho_\mathrm{s}s}{\rho_\mathrm{g}c_\mathrm{s}}\Omega_k \propto s \left(\dfrac{r}{r_0}\right)^p,
\end{equation}
this growth timescale $\tau_g$ at a given distance to the star $r$ becomes:

\begin{equation}
\tau_g \propto \left\{
                \begin{array}{cl}
                  \left(\dfrac{r}{r_0}\right)^{\frac{3}{2}(p+1)} & \mathrm{St} \ll 1, \\
                   \left(\dfrac{r}{r_0}\right)^{\frac{1}{2}(p+3)} & \mathrm{St} \gg 1. \\
                \end{array}
              \right.
\end{equation}
The typical growth timescale is thus always an increasing function of the distance to the star.

\bsp	
\label{lastpage}
\end{document}